\documentclass[12pt]{amsart}
\usepackage{amsmath,amssymb,enumerate,epsf,epsfig,amsfonts,graphicx,color}
\usepackage[applemac]{inputenc}
\usepackage{hhline}
\usepackage{a4wide,epsfig}
\numberwithin{equation}{section}
\newtheorem{theorem}{Theorem}

\newcommand{\comm}[1]{}

\newcommand\op[1]{\mathop{\rm #1}\nolimits}

\newcommand\R{{\mathbb R}}

\def\R{\mathord{\mathbb R}}

\def\2{\frac{1}{2}}
\def\3{{\ss}}

\def\.{\cdot}

\def\<{\langle}
\def\>{\rangle}

\def\beq{\begin{equation}}
\def\eeq{\end{equation}}
\def\bea{\begin{eqnarray}}
\def\eea{\end{eqnarray}}
\def\bsm{\left(\begin{smallmatrix}}
\def\esm{\end{smallmatrix}\right)}
\def\bpm{\begin{pmatrix}}
\def\epm{\end{pmatrix}}

\newcommand{\weg}[1]{}

\date{}
 \title[Nonexistence of the  6th degree integral for  Zipoy-Voorhees metric]{Nonexistence 
of an integral of the 6th degree  in momenta for the Zipoy-Voorhees metric}
 \author{Boris S. Kruglikov and  Vladimir S. Matveev}
 \address{B.\,S. Kruglikov: Department of Mathematics and Statistics,
University of Troms\o, 9037 Troms\o, Norway.
Email: {\tt boris.kruglikov@uit.no}}
 \address{V.\,S. Matveev: Institute of Mathematics, Friedrich-Schiller-Universit\"at, 07737 Jena, Germany.
Email: {\tt vladimir.matveev@uni-jena.de}}
 
\begin{document}
\maketitle

{\bf Abstract:} We prove nonexistence of a nontrivial  integral  that is polynomial  in momenta of degree $\le$6
for the Zipoy-Voorhees spacetime with the parameter $\delta=2$.

 \section{Introduction and the result}
 
We  consider the following 4-dimensional metric on an open set $U\subset \mathbb{R}^4$:
 \begin{equation}\label{g} 
ds^2 = \textrm{\tiny
$\left(\frac{x +1}{x-1}\right)^2 \left[ (x^2 -y^2)\left(\frac{x^2 -1}{x^2-y^2}\right)^4 \left(\frac{dx^2}{x^2 -1 } + \frac{dy^2}{1-y^2}\right) + (x^2 -1)(1-y^2)d\phi^2\right]
-\left(\frac{x-1}{x+1}\right)^2dt^2. $}
 \end{equation}

% Here $(x,y,\phi,t)$ are local coordinates on $U$.
% $(x,y,\phi,t)=(x_1,x_2, x_3, x_4)$

This metric was introduced  in \cite{Zipoy,Voorhees}.  
It is Ricci-flat, has Lorenz signature, and two commuting Killing  vector fields: 
$\partial_t$ ($ds^2$ is static) and $\partial_\phi$ ($ds^2$ is axially-symmetric).

Our interest in studying this metric is motivated by the resent  papers  of J. Brink
\cite{Brink1,Brink2,Brink3,Brink4}. She constructed numerically the geodesics of this metric and found out that they behave quite unusual. As it is clearly seen on figure 6 in \cite{Brink2}, the behavior of the geodesics is in no way chaotic and  is  typical for the behavior of trajectories of a Liouville-integrable system. Later, in a private conversation, Brink  demonstrated us  that this `integrable' behavior happens for all  geodesics.

Recall that a function $I:T^*U\to \mathbb{R}$ is called an {\it integral} for the geodesic flow of the metric $g$ 
if it {\it Poisson-commutes} with the kinetic energy $H=\tfrac12g^{ij}p_ip_j$.
Denoting by $\{\,,\}$ the canonical Poisson bracket this condition writes as $\{I, H\}=0$, and it means that 
the function $I$ is constant on the trajectories of the Hamiltonian system. In other words, $I$
is the conserved quantity of the Hamiltonian $H$. A metric $g$ is {\it Liouville-integrable\/} (for $\dim U=4$)
if, in addition to the Hamiltonian, there exist 3 Poisson-commuting functionally independent integrals 
for the geodesic flow. 
Recall that integrals are {\it functionally independent} if their differentials are linearly independent
almost everywhere.

Metric \eqref{g} already has two additional commutative integrals, namely the momenta $p_\phi$ and $p_t$; 
the numerical behavior of the geodesics is an indication of the existence of a fourth integral,
commuting with $p_t$ and $p_\phi$.

In the papers  \cite{Brink3,Brink4}, Brink  suggested to look for an integral of the geodesic flow that is
a homogeneous polynomial in momenta (with coefficients depending on the position). 
There are two  reasons for this choice of an ansatz for the integral. 
The first one is that for all physically interesting systems the known integrals are such.
The second reason is more mathematical: since the metric is real-analytic, it is expected that the integral 
is real-analytic as well. Now, as it was known already to Whittaker   \cite{Whittaker},
the existence of an integral analytic in momenta implies the existence of an integral that is a homogeneous
polynomial in momenta (More precisely, if $F= \sum _{i=0}^\infty F_k$ is an integral, where $F_k$ is a homogeneous in momenta of degree $k$, then every $F_k$ is an integral. If $F$ commutes with $p_t$ and $p_\phi$, then every $F_k$ commutes with $p_t$ and $p_\phi$). 
Thus we can restrict to the integrals, which are polynomials of degree $k$ in momenta. 
They are essentially the same objects as Killing $(0,k)$-tensors.

The main statement of  our paper is that, contrary to numerical observations, no  nontrivial
integral of the expected low degree exists. The `trivial' integrals for the geodesic flow of the metric $g$ 
are $H$, $I_1=p_\phi$, $I_2=p_t$ and any (polynomial) function of them.

 \begin{theorem}
There exists  no smooth  function $I:T^*U\to \mathbb{ R}$, which is a polynomial in momenta of
degree $\le 6$, such that $H$, $I_1$, $I_2$, $I$ are functionally independent and Poisson-commute.
 \end{theorem}

This is a rigorous mathematical statement and its proof is also rigorous, though it is heavily based on 
computer algebra calculations. In the next section we explain the mathematical foundations and the details
of these computations. The calculations  are too complicated to be presented  here in all the details, but the idea behind them is rather simple
and can be easily realized in most computer algebra packages (the Maple worksheet is available from the authors). 

These computations can be extended to decide existence of higher degree integrals --
the algorithm is unchanged (in fact, we have also verified nonexistence of the integral of degree 7, 
but the hardware capacity limited us in degree 8).
We will give some further comments in the conclusion.

 \section{Nonexistence of the integrals}

In this section we demonstrate nonexistence of a nontrivial integral of degree 6.
This in turn implies nonexistence of a nontrivial integral of smaller degree.
Indeed if such an integral $I$ existed in degree $k<6$, 
then $p_t^{6-k}I$ would be a nontrivial integral of degree 6.

\subsection{The general idea behind the calculations}\label{idea}
The condition that the function 
 \begin{equation}\label{I}
I= \sum_{i+ j + k+ m=6} I_{ijkm}p_x^ip_y^jp_\phi^k p_t^m
 \end{equation} 
commutes with $H$ is equivalent to a linear system $S$ of PDE of the first order
on the coefficients $I_{ijkm}$; the latter are smooth functions on $U$. 
Using the assumption that $I$ commutes with $p_{\phi}$ and $p_t$, we see that 
the unknowns $I_{ijkm}$ are functions of $x$ and $y$ only. 
More details on the differential equations in $S$ will be given in the next subsection. 

Let us now take this system $S$ and differentiate its equations with respect to the 
variables $x$ and $y$, denoting the results by $S_x$ and $S_y$ respectively.
The system $S^{(1)}=S\cup S_x\cup S_y$ is called {\it the first prolongation} of $S$. 
Every smooth solution of $S$ is of course a solution of $S^{(1)}$. 
Next we consider the second prolongation $S^{(2)}=S^{(1)}\cup S_{x^2} \cup S_{xy} \cup S_{y^2}$
and so on. The {\it $n$-th prolongation\/} is $S^{(n)}=\bigcup_{k+m\le n} S_{x^ky^m}$.

The system $S$ has {\it finite type} if for some $\ell$ one can solve
the equations $S^{(\ell)}$ with respect to all highest derivatives of the unknown functions
(in our case with respect to all jets of order $\ell+1$). 
It is known \cite{Wolf,kruglikov} that our system  is of finite type.
To see this let us notice that in the geodesic coordinates the metric is flat to
the second order at the given point. 
Since the coefficients of the top derivatives of the unknowns in the system $S$ 
(and its prolongation $S^{(n)}$) depend only on the zeroth and first derivatives of $H$,
its symbolic behavior at any point is the same as for the flat metric. In other words, 
the symbols of this system are isomorphic at all points and achieve finite type at the same level $\ell$.

The calculation for the flat metric is not difficult, and we obtain $\ell=k$ for the system describing 
integrals of degree $k$. In particular, for our $S$ we need to prolong $\ell=6$ times to achieve finite type
(a calculation supporting this claim will be shown in \S\ref{details}). 

For every linear system of finite type, the space of solutions is a finite dimensional vector space. 
Indeed, by the classical argument, the solutions are given as integral surfaces of the 
Cartan distribution, which in our case has rank 2 (on the equation-manifold given in the space
of $(\ell+1)$-jets by $S^{(\ell)}$).
Thus a solution $I$ of $S$ is uniquely determined by the values of its derivatives up to order $\ell+1$ 
(here 1 is the order of $S$) at any fixed point $(x_0,y_0)$.
This follows from the observation that restriction of $I$ to a curve $\gamma\subset\R^2(x,y)$
through the point $(x_0,y_0)$ reduces $S$ to a system of ODE in the Euler form.

Take $n\ge\ell$ and consider the prolonged system $S^{(n)}$. 
Treating the derivatives of $I_{ijkm}$ at $(x,y)$
up to the order $n+1$ as independent variables $u$ (called {\it jets\/}) 
allows us to write this linear system in the form
$Au=0$. Because coefficients of the Zipoy-Voorhees metric $ds^2$ are algebraic (rational) functions,
the entries of the matrix $A$ are also algebraic as functions of the point $(x,y)$.

If at a certain point $(x_0,y_0)$ the rank of the matrix $A$ equals to the dimension of $u$, 
the only solution is trivial $u=0$. From the finite type condition it follows then that 
the solution $u$ is identically zero.

As we explain in the next section, our PDE system $S$ decouples into two linear subsystems 
which have to be solved independently. For one of them $S_{odd}$ the situation is as described above: 
for the $5^\text{th}$ prolongation the linear system $Au=0$ has only the trivial solution.
For the second subsystem $S_{even}$ the situation is more complicated, since it has 
16-dimensional space of solutions. These solutions correspond to the following `trivial' integrals:

 \begin{equation} \label{trivial}
I_{\rm triv}= \alpha H^3 + H^2\cdot\sum_{i=0}^2\beta_i I_1^i I_2^{2-i}+ 
H\cdot\sum_{j=0}^4\nu_j I_1^j I_2^{4-j} + \sum_{k=0}^6\eta_k I_1^k I_2^{6-k},
 % \sum_{i=0}^q \eta_i p_\phi^i p_t^{q-i}
 \end{equation}
where $\alpha, \beta_i, \nu_j, \eta_k$ are  arbitrary constants.

Thus for $S_{even}$ the matrix $A$ has a 16-dimensional kernel, whence $rk(A)\le\dim(u)-16$.
If $rk(A)=\dim(u)-16$, then every solution $u$ of the equations $Au=0$ is `trivial' because
the system is of finite type and $n\ge\ell$. In other words, $u$ has form \eqref{trivial} globally. 

How do we check that the finite type level $\ell$ does not exceed the number $n$ we consider?
If for a certain $n$ we have $rk(A)=\dim(u)$, the condition $\ell\le n$ is fulfilled automatically
since we can express from the equations in $S^{(n)}$ all derivatives of the unknown functions.
This finishes the case $S_{odd}$. In fact in this case $\ell=5$ (this is a simpler system than
$S_{even}$ and it achieves finite type earlier) and we go to $n=5$ in our calculations.

For $S_{even}$ for $n=6$ we will have $\dim(u)-rk(A)=16$, which equals to the dimension 
of the space of trivial integrals of degree 6.
Since $\ell=6$ by the general theory, we conclude that
every integral must belong to the family \eqref{trivial}, which finishes our argument. 

The size of the matrix $A$ in our calculations is quite big, and it is not possible to handle it
by hand. We use the symbolic software Maple. This is possible as the 
idea described above can be realized algorithmically.

Actually we perform prolongations (symbolic differentiations) and operate with polynomial or rational functions. 
The coefficients of the latter are rational numbers and the point we substitute is also rational -- 
we choose $(x_0,y_0)=(\frac12,2)$. Then we calculate ranks of the matrices with rational numerical entries, 
which is done via the Gauss method. All of these calculations are exact (no approximations), and so the result 
has to be considered as a computer assisted mathematically rigorous proof.

\subsection{Details of calculations}\label{details}

We look for an integral $I$ of form (\ref{I}) with coefficients $I_{ijkm}$ being smooth functions on $U$. 
Since $\{I,I_1\}=\{I,I_2\}=0$, these coefficients do not depend on $\phi,t$ and are smooth functions
on the open domain in $\R^2(x,y)$ (given by the conditions $x,y,\ne\pm1$, $x\ne\pm y$).

Since the coefficients of $H$ also do not depend on $\phi,t$, the commutation of $H$ and $I$ writes as
 \begin{equation}\label{poisson}
\{H,I\}=\frac{\partial H}{\partial x}\frac{\partial I}{\partial p_x}-
\frac{\partial I}{\partial x}\frac{\partial H}{\partial p_x}+ 
\frac{\partial H}{\partial y}\frac{\partial I}{\partial p_y}-
\frac{\partial I}{\partial y}\frac{\partial H}{\partial p_y}=0.
 \end{equation}
Substituting formulae for $H$ and $I$ into this expression we get a  
polynomial in momenta of degree 7. Its vanishing is equivalent to vanishing of
all its coefficients, and there are 120 of them. These coefficients are rational functions of $(x,y)$
with integer coefficients and they are linear in $I_{ijkm}$ and their first derivatives. 
The number of unknowns $I_{ijkm}=I_{ijkm}(x,y)$ is 84, and thus 
\eqref{poisson} gives a first order linear overdetermined system $S$ of PDE.

Notice that the Hamiltonian $H$ is even in the total degree by the variables $p_\phi,p_t$
(i.e. it does not contain quadratic terms $p_xp_t$ and similar). Then writing $I=I_{odd}+I_{even}$,
where the index refers to the parity of $k+m$ in \eqref{I}, we observe the splitting
$\{H,I\}=\{H,I_{odd}\}+\{H,I_{even}\}$ and the summands in the last expression preserve
the parity (e.g. $\{H,I_{odd}\}$ has odd total degree by the variables $p_\phi,p_t$).

Thus our system of equations decouples: 
$S=S_{odd}\cup S_{even}$, $S_{odd}\cap S_{even}=\emptyset$. 
The subsystem $S_{odd}$ is a system of linear PDE on $I_{ijkm}$ with $k+m$ odd,
and the subsystem $S_{even}$ is a system of linear PDE on $I_{ijkm}$ with $k+m$ even.
Both $S_{odd}$ and $S_{even}$ are overdetermined.

In fact $S_{odd}$ consists of 60 equations on 40 unknown functions, while 
$S_{even}$ consists of 60 equations on 44 unknowns. Our goal for the subsystem $S_{odd}$ is to show that 
its certain prolongation at some point $(x_0,y_0)$ satisfies $\dim(u)-rk(A)=0$.  
Our goal for the subsystem $S_{even}$ is to show that its certain prolongation at 
$(x_0,y_0)$ satisfies $\dim(u)-rk(A)=16$.
We follow the scheme described in \S\ref{idea} and let $(x_0,y_0)=(\frac12,2)$.

Now we put the results of our calculations into the table.
The first row in the table is the number of equations in $S_{odd}^{(n)}$, 
the second is the number of the unknowns $I_{ijkm}$ (with $k+m$ odd) and their derivatives 
by $x,y$ up to order $n+1$. The third row is the rank of the corresponding matrix $A$ 
of the size (\# of eqn)$\times$($\dim(u)$).

The number of nontrivial integrals is the minimum of the quantity $\Delta=\dim(u)-\op{rk}(A)$.

\begin{tabular}{c |c|c|c|c|c|c}
  $n$            & 0   & 1   & 2   & 3   & 4   & 5    \\ \hline
  {\# of eqn}    & 60  & 180 & 360 & 600 & 900 & 1680 \\ \hline
  {$\dim(u)$}    & 120 & 240 & 400 & 600 & 840 & 1440 \\ \hline
  {$\op{rk}(A)$} & 60  & 180 & 360 & 590 & 838 & 1440 \\ \hline
  {$\Delta$}     & 60  & 60  & 40  & 10  & 2   & 0
\end{tabular}

We see that the $5^\text{th}$ prolongation is enough to prove that the system $S_{odd}$ has
only trivial solutions.

Next comes the table for $S_{even}$. The meaning of the rows is similar.
The number of nontrivial integrals is the minimum of the quantity $\Delta=\dim(u)-\op{rk}(A)-16$.

\begin{tabular}{c |c|c|c|c|c|c|c}
  $n$            & 0   & 1   & 2   & 3   & 4   & 5    & 6    \\ \hline
  {\# of eqn}    & 60  & 180 & 360 & 600 & 900 & 1260 & 1680 \\ \hline
  {$\dim(u)$}    & 132 & 264 & 440 & 660 & 924 & 1232 & 1584 \\ \hline
  {$\op{rk}(A)$} & 60  & 180 & 360 & 600 & 888 & 1215 & 1568 \\ \hline
  {$\Delta$}     & 56  & 68  & 64  & 44  & 20  & 1    & 0
\end{tabular}

We see that the $6^\text{th}$ prolongation is enough to prove that the system $S_{even}$ has only 
'trivial' solutions, i.e. every integrals $I_{even}$ is of the form \eqref{trivial}.

Finally we demonstrate how to calculate the number $\ell$ symbolically.
As explained in \S\ref{idea} it is enough to compute it for the flat metric.
We form equation $S$ similar to the above, but now we count only
the equations of order $n+1$ in $S^{(n)}$, and we separate top-order terms.
The latter are the derivatives of $I_{ijkm}$ of order $n+1$ denoted by $v$,
and the subsystem writes as a linear inhomogeneous equation $Bv=w$, where $w$ 
combines derivatives of order $\le n$. When $\Delta=\dim(v)-rk(B)=0$ the corresponding  
$n=\ell$. Here is the table for $S_{even}$.

\begin{tabular}{c |c|c|c|c|c|c|c}
  $n$            & 0  & 1   & 2   & 3   & 4   & 5   & 6   \\ \hline
  {\# of eqn}    & 60 & 113 & 166 & 219 & 272 & 325 & 378 \\ \hline
  {$\dim(v)$}    & 88 & 132 & 176 & 220 & 264 & 308 & 352 \\ \hline
  {$\op{rk}(B)$} & 60 & 113 & 166 & 214 & 262 & 307 & 352 \\ \hline
  {$\Delta$}     & 28 &  19 &  10 &  6  &  2  &  1  &  0
\end{tabular}

Thus $\ell=6$. Similar calculations for $S_{odd}$ yield $\ell=5$. This finishes the proof of the theorem.

 \section{Conclusion}

The question of existence of an additional integral is crucial for understanding 
of static axially-symmetric Ricci-flat metrics.
The  most famous spacetimes in the general relativity, given by the Schwarzschild and Kerr metrics, 
can be effectively studied because their geodesic flows admit an additional quadratic integral.
This integral allows to describe and to control the behavior of the geodesics, 
and it also helps solving the wave and other physically-relevant equations.

The family of Zipoy-Voorhees metrics with a real parameter $\delta$ is given by the formula
 $$
\Bigl(\frac{x+1}{x-1}\Bigr)^\delta\Bigl((x^2-y^2)\Bigl(\frac{x^2-1}{x^2-y^2}\Bigr)^{\delta^2}
\Bigl(\frac{dx\,dx}{x^2-1}+\frac{dy\,dy}{1-y^2}\Bigr)+(x^2-1)(1-y^2)\,dz\,dz\Bigl)-
\Bigr(\frac{x-1}{x+1}\Bigr)^\delta dt\,dt.
 $$
Transformation $(x,\delta)\mapsto(-x,-\delta)$ is the symmetry of this family, so we let $\delta\ge0$.
Parameter $\delta=0$ corresponds to the flat metric, while for $\delta=1$ we get the Schwarzschild metric.
These are the only two cases from the family that admit Killing vector fields. For the Schwarzschild 
spacetime the symmetry algebra is in fact non-commutative, but its nontrivial Casimir function is 
an additional integral that yields Liouville integrability.

We have checked that no other metric from the family admits a quadratic integral. The search for higher 
degree integrals we restricted to the next interesting parameter $\delta=2$, corresponding to
the metric \eqref{g}. We hoped that the integral of degree 4 exists; some other groups of mathematicians 
have also tried to find the missing integral by looking for a lucky ansatz.
But, unfortunately, we have not found it and moreover we have proved it cannot exist in degree $<7$. 
Degree 7 we have investigated with other methods (not the one presented in the paper), which also
yield nonexistence. 

We stopped at degree 8 by purely technical reasons -- even though the operational memory was not 
suffering, the computation time got too long. Actually, Maple obtains the matrix $A$ quite fast; 
the hard part of the calculations is to compute the rank of the matrix $A$, which for the integrals 
of degree 6, has size $1680\times 1584$. Our standard PC required few days for this. But we hope 
that a faster computer and more specialized symbolic software (which handles rational numbers in
exact manner) could advance further.

Let us remark that the Zipoy-Voorhees metrics can be included in the bigger family of Manko-Novikov metrics, 
which are also given by explicit formulas. Every axially-symmetric stationary Ricci-flat metric can be written 
with the help of the Ernst equations. Having explicit formula for the integral for the Zipoy-Voorhees 
spacetime would suggest a perturbaton of this formula to integrate other stationary 
axially-symmetric Ricci-flat metrics.
J. Brinck calculated numerically also the geodesics of certain Manko-Novikov metrics 
and indicated numerically an integrable behavior.

Existence of an integral of high degree is not completely impossible, as recently 
some nontrivial superintegrable systems were found. Its search can follow the proposed algorithm.
One can further simplify it as it is apparent that the matrix $A$ contains a lot of zeros.
Another possibility is to use geometric methods to investigate the structures generated by the integral  
(in this way \cite{degree3} allowed to find a new integrable system in \cite{dullin}). 

Anyway the advantages of having an integral are so huge that the search for additional integrals 
of axially symmetric Ricci-flat metrics should be continued. But as our results indicate 
if the spacetime \eqref{g} is integrable, it should be rather a nontrivial phenomenon.

\subsection*{Acknowledgements.}  We thank J. Brink and R. Meinel  for useful discussions and   the universities of Jena and Troms\o {} for hospitality.  This research was partially supported by the
 DAAD exchange project Nr. 208068.


\begin{thebibliography}{100}

\bibitem{degree3} H.\,R. Dullin,  V.\,S. Matveev, P. Topalov,
{\it On integrals of third degree in momenta\/},
Regular and Chaotic Dynamics, {\bf 4} (1999), 35--44.

\bibitem{dullin} H.\,R. Dullin,  V.\,S. Matveev,
{\it New integrable system on the sphere\/},
Math. Res. Lett. {\bf 11} (2004), 715--722; arXiv:0406209.

\bibitem{Zipoy} D.\,M. Zipoy, 
{\it Topology of some spheroidal metrics\/},
J. Math. Phys. (N.Y.) {\bf 7} (1966), 1137ñ-1143.

\bibitem{Voorhees} B. Voorhees, 
{\it Static Axially Symmetric Gravitational Fields\/},
Phys. Rev. D {\bf 2} (1970), 2119.

\bibitem{Brink1} J. Brink,
{\it Spacetime encodings. I. A spacetime reconstruction problem\/},
Phys. Rev. D {\bf 78} (2008), 102001.

\bibitem{Brink2} J. Brink,
{\it Spacetime encodings. II. Pictures of integrability\/},
Phys. Rev. D {\bf 78} (2008), 102002.

\bibitem{Brink3} J. Brink,
{\it Spacetime encodings. III. Second order Killing tensors\/},
Phys. Rev. D {\bf 81} (2010), 022001.

\bibitem{Brink4} J. Brink,
{\it Spacetime encodings. IV. The relationship between Weyl curvature and Killing tensors in stationary axisymmetric vacuum spacetimes\/},
Phys. Rev. D {\bf 81} (2010), 022002.

\bibitem{kruglikov} B. Kruglikov, 
{\it Invariant characterization of Liouville metrics and polynomial integrals\/},  
J. Geom. Phys. {\bf 58} (2008),  no. 8, 979--995; arXiv:0709.0423.

\bibitem{Wolf} Th. Wolf, 
{\it Structural equations for Killing tensors of arbitrary rank\/},
Comput. Phys. Comm. {\bf 115} (1998), no. 2-3, 316--329.

\bibitem{Whittaker} E.\,T. Whittaker, 
{\it A treatrise on the Analytical Dynamics of Particles and Rigid Bodies\/}, 
Cambridge University Press (1937).

\end{thebibliography}
\end{document}